\documentclass[]{main}

\usepackage{array}
\usepackage{epstopdf}
\usepackage[caption=false]{subfig}
\usepackage[doublespacing]{setspace}
\usepackage[nodoi]{apacite}

\begin{document}

\title{InSight: A Mobile Assistant Helping Visually Impaired People Navigate Unfamiliar Environments}

\author{
\name{Junwei Zheng\textsuperscript{a}, Jiale Wei\textsuperscript{a}, Di Wen\textsuperscript{a}, Yufan Chen\textsuperscript{a}, Ruiping Liu\textsuperscript{a}, Chengzhi Wu\textsuperscript{a}, Kunyu Peng\textsuperscript{a}, Kailun Yang\textsuperscript{b} and Jiaming Zhang\textsuperscript{b,*}\thanks{*Corresponding author: Jiaming Zhang. Email: jiamingzhang@hnu.edu.cn}}
\affil{\textsuperscript{a}Karlsruhe Institute of Technology, Karlsruhe, Germany;  \\
\textsuperscript{b}Hunan University, Changsha, China}
}

\maketitle

\begin{abstract}
Navigating unfamiliar outdoor environments is one of the most demanding everyday tasks for people with visual impairments (PVI).
However, existing navigation systems cannot simultaneously satisfy the needs of informative perception and precise localization during navigation.
This paper explores the core navigation demands of PVI through a preliminary study.
Based on the results of the pre-study, we present the design of \textit{InSight}, a mobile assistant that combines scene descriptions with precise localization to support PVI navigating unfamiliar outdoor environments.
The offline mode of \textit{InSight} runs entirely on device for real-time navigation, while the online mode streams camera frames to a self-hosted server for arbitrary object detection and accurate localization.
Finally, a user study with 21 visually impaired participants validates that \textit{InSight} could assist users in navigating unfamiliar outdoor environments with increased success rate and confidence compared to mainstream navigation applications like Apple Maps.
\end{abstract}

\begin{keywords}
Visual impairment; assistive technology; pedestrian navigation; scene understanding; accurate localization
\end{keywords}

\section{Introduction}\label{sec:intro}

Navigating independently to a destination in an unfamiliar outdoor environment is a substantial challenge for people with visual impairments (PVI).
With their regular navigation aids such as the white cane or a guide dog, safe navigation relies on long-term familiarization.
PVI build cognitive maps of an area from tactile and auditory information~\cite{Afonso2022SpatialAudio,ottink2022cognitivemap}.
In an unfamiliar outdoor environment, familiarity does not exist, while moving traffic, temporary obstacles, and complex intersections require decisions in real time.
PVI therefore often navigate unfamiliar destinations accompanied by sighted people, or must rely on passersby for directions and on-site help when traveling alone~\cite{muller2022traveling}.
Despite these difficulties, traveling to unfamiliar places remains a part of daily life, and PVI express a strong desire to complete such travel by themselves rather than relying on sighted assistance~\cite{jain2023want,muller2022traveling}.
These observations underscore the need for navigation assistance that enables PVI to reach destinations in unfamiliar outdoor environments independently.

However, existing navigation technologies do not meet the requirements.
Mainstream navigation applications such as Apple Maps and Google Maps offer limited assistance, as they are designed for sighted users and only provide PVI-unfriendly turn-by-turn instructions with inaccurate positioning.
Their positioning inherits the errors of the Global Positioning System (GPS), which typically amount to several meters and can grow to tens of meters in urban canyons~\cite{zangenehnejad2021gnss}.
Sighted pedestrians compensate for such errors visually, whereas for PVI the same errors can place an instruction on the wrong side of an intersection.
Accessible navigation applications designed for PVI, such as BlindSquare and Microsoft Soundscape, convey nearby points of interest and street information through audio~\cite{soundscape2018,blindsquare}.
Since they rely entirely on pre-built map data and the same imprecise GPS positioning, these applications cannot inform users about real-world obstacles absent from the map, such as parked vehicles, pedestrians, or construction sites.
Assistive applications that provide visual descriptions, such as Seeing AI or Be My AI, require the user to stop, point the camera, and issue a request, which interrupts travel and occupies the hands~\cite{BeMyAI2023,seeingAI}.
In addition, remote sighted assistance services depend on the availability of another person~\cite{aira}.

In this paper, we aim to design a mobile navigation assistant that enables PVI to navigate unfamiliar outdoor environments independently by addressing the following research questions (RQ):
\begin{enumerate}
\item[\textbf{RQ1:}] What assistive features do PVI need when navigating unfamiliar outdoor environments?
\item[\textbf{RQ2:}] To what extent can PVI navigate unfamiliar outdoor environments independently with the support of a navigation assistant?
\item[\textbf{RQ3:}] How does navigation assistance affect the perceived workload and user experience of PVI in unfamiliar environments?
\end{enumerate}

To answer RQ1, we conducted a formative preliminary study.
We first observed a visually impaired person navigating outdoors.
We then interviewed him about which assistive features matter during navigation.
Finally, we co-designed an online questionnaire with him and collected responses from five PVI.
Based on the findings, we designed \textit{InSight}, a mobile application that supports outdoor navigation for PVI with two complementary functions.
First, \textit{InSight} provides arbitrary object detection.
The phone camera continuously detects surrounding objects, not limited to a predefined set of categories.
For each detected object, \textit{InSight} reports its category, distance, and direction, for example, ``bicycle in two meters at eleven o'clock.''
Second, \textit{InSight} provides precise localization.
It aligns camera observations with OpenStreetMap to correct the user's position and heading.
Along the planned route, \textit{InSight} then conveys upcoming turns as clock directions, for example, ``turn to two o'clock.''
Both functions express directions in the same clock-face format, providing the user with a single mental model.
\textit{InSight} operates in two modes.
The offline mode runs entirely on the phone in real time, whereas the online mode streams camera frames to a self-hosted server, where both detection and localization are enhanced by advanced algorithms.

To answer RQ2 and RQ3, we evaluated \textit{InSight} in a summative user study with 21 visually impaired participants.
Each participant navigated unfamiliar outdoor routes under three conditions: Apple Maps as a commercial baseline, the offline mode of \textit{InSight}, and its online mode.
Compared with Apple Maps, both modes led to a higher navigation success rate, a lower perceived workload, and better usability ratings, with the online mode performing best.

We summarize the contributions of this paper as follows:
\begin{itemize}
\item We present \textit{InSight}, a navigation assistant for PVI pedestrians that integrates precise localization and arbitrary object detection in one system, with a fully on-device offline mode and a server-enhanced online mode.
\item We provide empirical evidence from the complete research lifecycle. A formative pre-study identifies what PVI need in unfamiliar outdoor navigation (RQ1), and a summative within-subject field study with 21 PVI participants measures the extent to which those needs are met (RQ2, RQ3).
\item Based on the findings, we discuss the capabilities and limitations of the system and derive design implications for future navigation assistants for PVI.
\end{itemize}

\section{Related Work}\label{sec:related}

\subsection{Navigation Assistance for People with Visual Impairments}\label{sec:navigation}

Different from sighted pedestrians, PVI sense the path with the cane and rely on tactile or auditory landmarks rather than visual information~\cite{williams2014just}.
In unfamiliar environments where such landmarks are unknown, disorientation occurs easily~\cite{alkhanifer2015disorientation,jeamwatthanachai2019indoor}.
Existing works have therefore developed assistive navigation systems for PVI.
Microsoft Soundscape reports surrounding streets and points of interest through spatialized audio and marks the destination with an audio beacon~\cite{soundscape2018}.
Related systems use spatial audio to convey nearby places~\cite{blum2012what} or to guide the user along a route~\cite{katz2012navig}.
Other studies turn to dedicated hardware.
Wearable, handheld, or smartphone-based devices report direction through haptic cues~\cite{amemiya2009haptic,ando2012haptic,bouteraa2021wearable,liu2021tactile,pielot2012pocketnavigator}.
LightGuide instead directs the user along a path with light cues~\cite{yang2021lightguide}.
Sonic alarms clear paths through crowds~\cite{kayukawa2019bbeep}.
Drones or robots physically lead the user~\cite{kuribayashi2025wanderguide,liu2026not,tan2021flying}.
Beyond these aids, remote sighted assistance connects the user to a human guide through a video call, which offers flexible help but depends on the availability of another person~\cite{kamikubo2020support,lee2020emerging,lee2022opportunity}.
PathFinder~\cite{masaki2023pathfinder} navigates blind users through unfamiliar buildings without prebuilt maps, using a suitcase-shaped robot that detects intersections and reads signs.
Its study with seven blind participants showed that they navigated unfamiliar buildings with more confidence than with their regular aid.
However, the approach depends on a wheeled robot and was designed for indoor environments.
Carrying such a robot everywhere is inconvenient.
These works reveal a common trade-off.
Smartphone applications rely on map data and cannot describe the scene in real time.
Better support depends on dedicated devices or human assistance.
Automatic scene descriptions on an ordinary phone during outdoor navigation therefore remain underexplored.
\textit{InSight} fills this gap.
It runs on a standard smartphone without dedicated hardware.
While the user walks, it delivers route guidance together with descriptions of the surroundings.

\subsection{Visual Scene Understanding in Assistive Systems}\label{sec:scene_understanding}

Visual scene understanding has been widely used to help PVI perceive their surroundings.
Object detectors on mobile or wearable cameras report obstacles and everyday objects~\cite{liu2023realtime,Sugashini2024yologlass}.
Semantic segmentation further captures walkable areas, terrain, transparent obstacles, and material properties~\cite{ou2022indoor,yang2018unifying,zhang2021trans4trans,zheng2023materobot,zou2023realtime_passable}.
Panoramic or wide field-of-view perception broadens the spatial context observable in a single frame~\cite{fan2026more,hu2024deformable}.
Wearable prototypes couple such perception with spatial audio or multi-sensory feedback~\cite{hu2022stereopilot,li2023multi}.
Commercial applications such as Seeing AI, Envision, and Be My AI generate scene descriptions upon explicit user request~\cite{BeMyAI2023,envision2024,seeingAI,xie2024emerging}.
Recent research prototypes further apply large vision-language models, ranging from on-demand visual querying on a wearable~\cite{mathis2025lifeinsight} to context-aware live descriptions~\cite{chang2024worldscribe,zhang2025navigpt}.
These models have also been evaluated on assistive tasks such as scene description and spatial reasoning~\cite{jiang2025bench,li2026exploring}.
Despite this progress, a gap still remains for navigation.
Most detection-based systems are constrained to a closed vocabulary, typically the 80 categories of the COCO benchmark~\cite{lin2014microsoft}.
As a result, objects and hazards critical to PVI pedestrians, such as barrier posts, foldout signs, and construction sites, remain undetected~\cite{feng2026robust,islam2024crucial}.
Open-vocabulary models lift this restriction~\cite{kirillov2023segment,liu2023opensu,zheng2024ops} and have recently been adopted in assistive systems, for example, ObjectFinder for interactive object search~\cite{liu2026objectfinder}.
However, these systems require either intensive computational resources or dedicated hardware instead of running on standard mobile phones.
\textit{InSight} closes this gap.
Its offline mode runs a lightweight detector directly on the smartphone and reports surrounding objects continuously while the user walks.
Its online mode offloads open-vocabulary recognition to a self-hosted server, enabling on-device arbitrary object detection during navigation.

\subsection{Localizing for Pedestrian Navigation}\label{sec:localization}

Navigation assistance depends on accurate estimates of the user's position and heading.
Consumer GNSS positioning on smartphones typically achieves an accuracy of several meters in open areas and degrades further in urban canyons~\cite{zangenehnejad2021gnss}.
To obtain higher accuracy, many assistive systems rely on installed infrastructure such as Bluetooth beacons or on prebuilt maps~\cite{masaki2023pathfinder}.
However, equipping or mapping every outdoor area in advance is impractical.
Vision-based localization avoids physical infrastructure.
Camera-captured images are matched against pre-captured visual databases or reconstructed models~\cite{duh2020veye,jain2023towards,zeng2026faster,zheng2025scene}.
Cross-view geo-localization extends this idea by matching ground-level images against aerial imagery or map data, and advanced methods reach metric accuracy~\cite{hu2018cvmnet,zeng2026x,zheng2026seqloc}.
Panoramic bird's-eye-view mapping turns street-level views into map-style semantic representations~\cite{wei2024onebev,wei2026onebev++}.
In particular, RHO matches camera images against OpenStreetMap (OSM) data and returns a metric position together with the heading of the camera~\cite{zheng2026rho}.
Since OSM is free and available worldwide, this approach works without installing beacons or mapping each area in advance, which makes it attractive for assistive use.
However, to the best of our knowledge, no prior work has studied whether and how such vision-based positioning changes the navigation experience of PVI pedestrians.
Therefore, in this paper, we developed \textit{InSight} as a research probe to investigate how integrated scene understanding and precise positioning shape independent navigation of PVI in unfamiliar outdoor environments.

\section{Preliminary Study}\label{sec:prestudy}

Prior work has revealed a mismatch between the assistive features emphasized by researchers and those desired by PVI~\cite{bhanuka2023want}.
To answer RQ1, we therefore conducted a formative pre-study before designing the system.
The pre-study consisted of three parts.
A field observation session showed us how a visually impaired person navigates outdoors.
An interview with him identified the assistive features that matter during navigation.
An online questionnaire then collected importance ratings from five PVI.
The session and the interview were conducted first to ground the questionnaire in the needs of PVI rather than in researcher assumptions.
The findings of the whole preliminary study led to four design goals that guide the design of \textit{InSight}.

\subsection{Method}\label{sec:prestudy_method}

\begin{figure}[t!]
\centering
\includegraphics[width=0.85\textwidth]{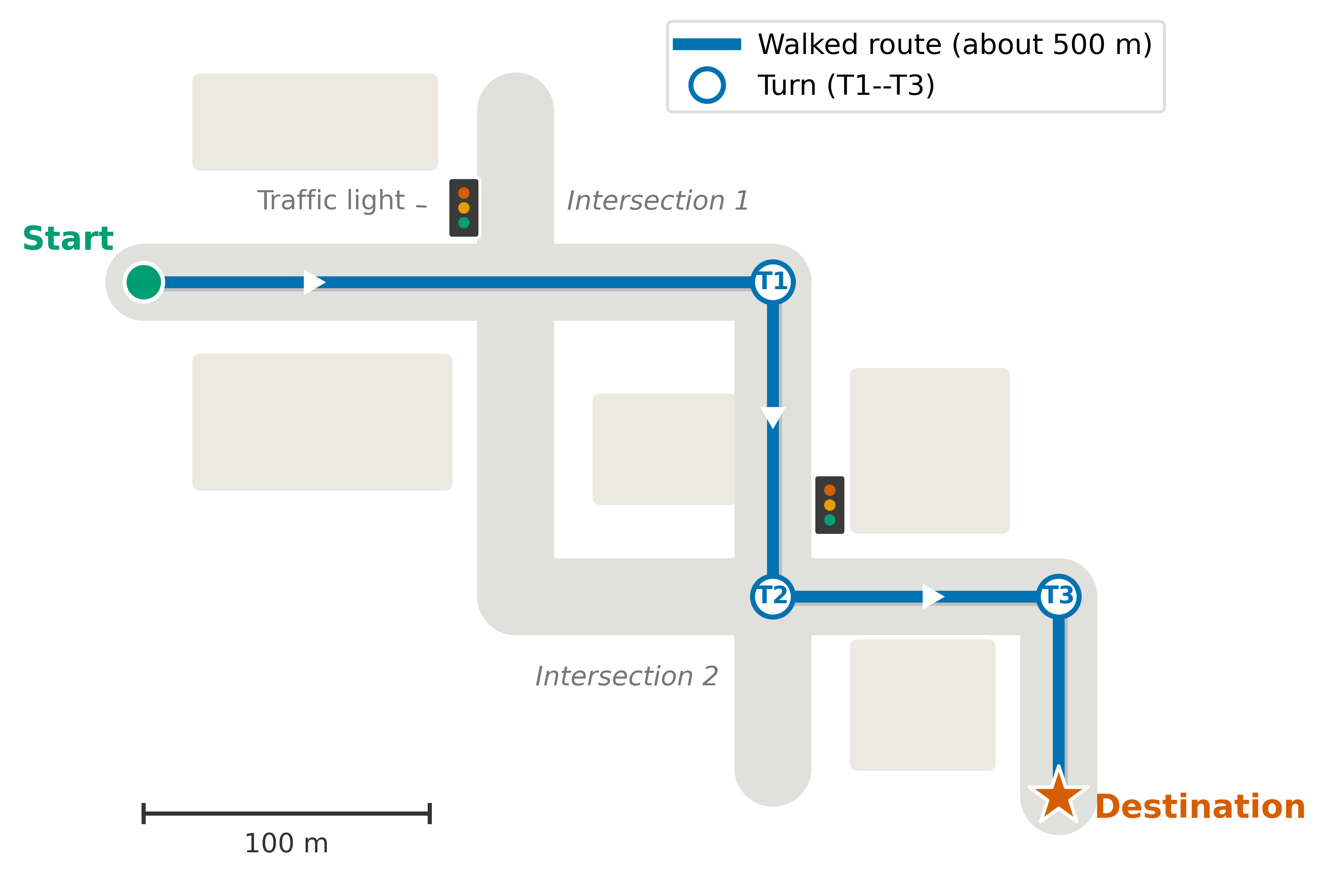}
\caption{The route of the field observation session. The co-designer walked about 500 meters from the start to the destination, passing two intersections with traffic lights and making three turns (T1--T3).}\label{fig:route}
\end{figure}

We first conducted a field observation session, in which we accompanied a visually impaired person as he navigated outdoors with BlindSquare.
He walked to a destination that he had not visited before, and the session lasted about 30 minutes.
The route was about 500 meters long and included two intersections with traffic lights and three turns.
Figure~\ref{fig:route} shows the route.
He was encouraged to think aloud during the walk.
We followed him closely for safety and took notes on how he used the application and where he met difficulties.
We then interviewed him to understand which assistive features matter during navigation in unfamiliar outdoor environments.
The upper part of Table~\ref{tab:prestudy} summarizes his profile.
With the help of the co-designer, we then designed an anonymous online questionnaire.
The field observation and the interview informed the selection of the candidate features, and the co-design refined the item wording and the response formats, for example, presenting each feature as a separate question with a fully labeled scale rather than in a matrix, which is difficult to navigate with a screen reader.
The questionnaire first collected demographic information, such as age group, gender, vision level, mobility aids and so on.
The main section asked respondents to rate the importance of candidate features on a 7-point scale, followed by optional open comments.
We collected valid responses from five PVI.
Their profiles are summarized in the lower part of Table~\ref{tab:prestudy}.

\begin{table}[t!]
\tbl{Demographics of the pre-study co-designer (C1) and the five questionnaire respondents (R1--R5).
Blind: a visual acuity of 2\% or less.
Severe impairment: above 2\% up to 5\%.
Low vision: above 5\% up to 30\%.}
{\setlength{\tabcolsep}{4pt}
\begin{tabular*}{\textwidth}{@{\extracolsep{\fill}}lllllll@{}} \\[-5.5pt]\noalign{\hrule height .8pt}\\[-5pt]
ID & Gender & Age group & Vision level & Onset & Mobility aids & Assistive apps \\ \midrule
C1 & Male & 50--59 & Blind & Since birth & White cane & Be My AI, BlindSquare \\ \midrule
R1 & Male & 20--29 & Blind & Since birth & White cane & Google Maps, Be My AI \\
R2 & Female & 30--39 & Blind & Later in life & White cane & Apple Maps, BlindSquare \\
R3 & Male & 40--49 & Low vision & Later in life & White cane & Seeing AI \\
R4 & Female & 20--29 & Blind & Since birth & White cane & Soundscape, Be My AI \\
R5 & Male & 50--59 & Severe impairment & Later in life & White cane & Google Maps, Be My AI \\ \\[-7.5pt]\noalign{\hrule height .8pt}
\end{tabular*}}
\label{tab:prestudy}
\end{table}

\subsection{Findings}\label{sec:findings}

\begin{table}[t!]
\tbl{The 15 candidate features of the pre-study questionnaire, grouped into six categories.}
{\begin{tabular*}{\textwidth}{@{\extracolsep{\fill}}lll} \\[-5.5pt]\noalign{\hrule height .8pt}\\[-5pt]
ID & Category & Feature \\ \midrule
F1 & Surrounding awareness & Descriptions of surrounding objects and obstacles \\
F2 & Surrounding awareness & Distance to obstacles \\
F3 & Surrounding awareness & Direction of obstacles \\
F4 & Localization & Exact position even where GPS is inaccurate \\
F5 & Localization & Exact facing direction even where the compass is inaccurate \\
F6 & Guidance & Step-by-step route instructions \\
F7 & Guidance & Confirmation of passed landmarks \\
F8 & Information delivery & Only necessary information \\
F9 & Information delivery & Control over the amount and frequency of speech output \\
F10 & Information delivery & No masking of environmental sounds \\
F11 & Information delivery & Clock-face directions \\
F12 & Privacy and practicality & No dedicated assistive device to carry \\
F13 & Privacy and practicality & On-device image processing \\
F14 & Extended assistance & Information about public transport \\
F15 & Extended assistance & Video calls to a sighted assistant \\ \\[-7.5pt]\noalign{\hrule height .8pt}
\end{tabular*}}
\label{tab:features}
\end{table}

\begin{figure}[t!]
\centering
\includegraphics[width=0.95\textwidth]{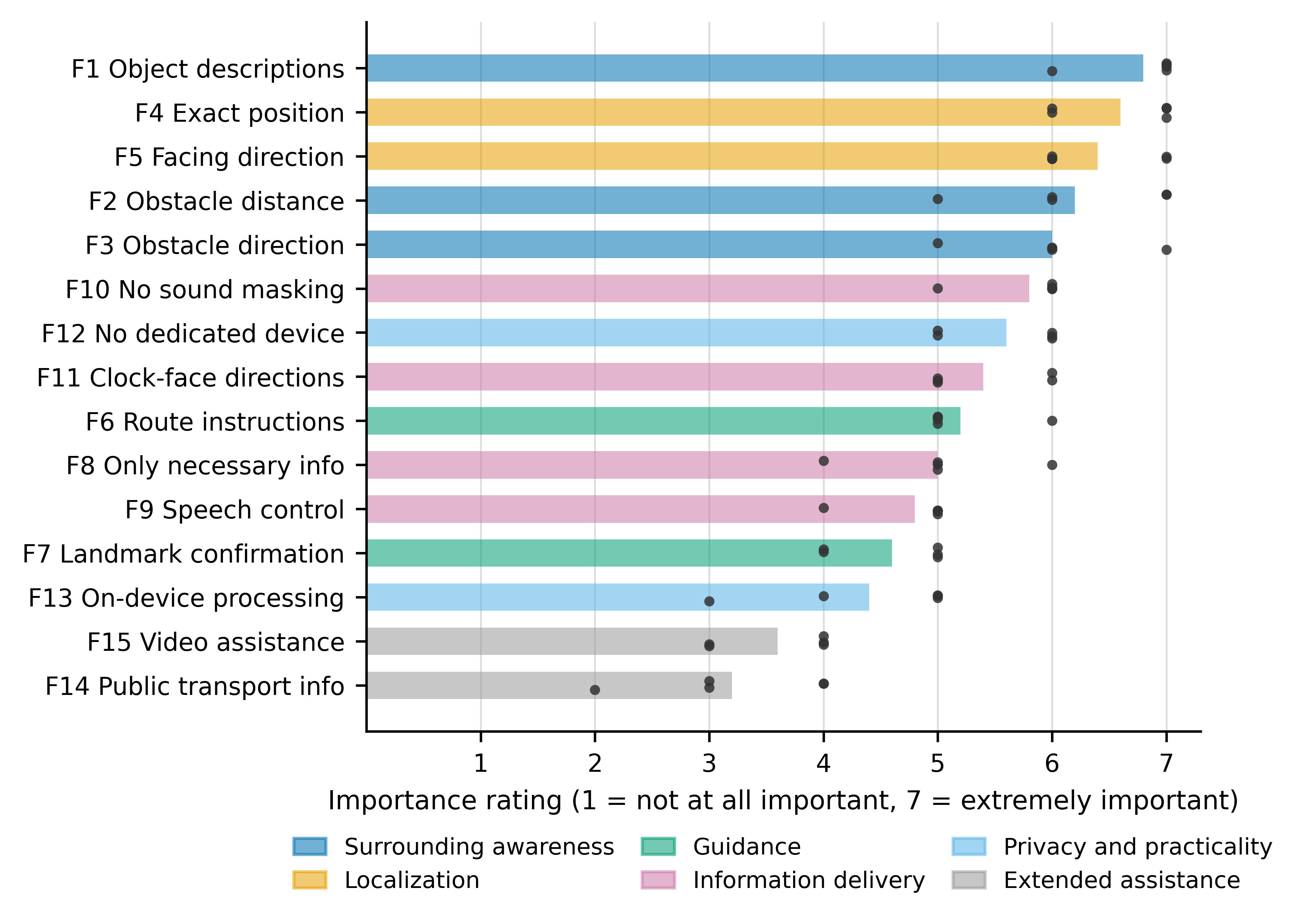}
\caption{Importance ratings (1 = not at all important, 7 = extremely important) of the 15 candidate features in the pre-study, ordered by mean rating. Bars show mean ratings, dots show the individual ratings of the five respondents, and colors indicate the six feature categories.}\label{fig:prestudy}
\end{figure}

The field observation confirmed the difficulties of navigating an unfamiliar route with current tools.
The co-designer walked slowly and stopped often to confirm his position and facing direction.
BlindSquare announced nearby streets and places, but its GPS-based position was not precise enough to tell him when to turn.
Thinking aloud, he repeatedly asked whether he had already reached the next crossing, and after one turn said that he no longer knew which direction he was facing.
The application also gave no information about obstacles along the way, which he had to probe with the white cane.
He bumped into an e-scooter that the cane had missed and nearly fell.
We stepped in at once to help him.
He said that such carelessly parked e-scooters were what he feared most.

In the interview, the co-designer talked about his everyday travel.
He said that he usually prepares unfamiliar routes in advance, for example by walking them first with a sighted friend.
He used BlindSquare for street information and Be My AI for scene descriptions.
However, neither helped him much while walking.
BlindSquare was not precise enough, and Be My AI required him to stop and hold up the phone.
He wished for an assistant that reports the surroundings and his exact position on the move, without occupying his hands or ears.

We translated the needs identified in the field observation and the interview into 15 candidate features.
Table~\ref{tab:features} lists these features in six categories: surrounding awareness, localization, guidance, information delivery, privacy and practicality, and extended assistance.
Figure~\ref{fig:prestudy} shows the importance ratings of each feature.
Features related to surrounding awareness and localization received the highest ratings.
In contrast, the two extended assistance features were rated lowest.

Respondents reported lower confidence when navigating unfamiliar places than familiar ones.
The most frequent challenges in unfamiliar places were knowing their exact position and avoiding unexpected obstacles.
These difficulties match the experiences that the co-designer described in the interview.
For example, he had been led to the wrong side of a crossing by inaccurate GPS positioning and had to ask passersby for help.

Regarding information delivery, most respondents preferred clock directions over left and right instructions and wanted automatic announcements only for important objects.
Attitudes toward processing camera images on a server were mixed.
Some respondents accepted server processing in exchange for better system performance.
Others preferred fully offline operation.

\subsection{Design Goals}\label{sec:goals}

The field observation, the interview, and the questionnaire point to the same needs.
PVI want more than route instructions.
They want to know what is around them and where exactly they are.
They also want to control how this information reaches them.
The eight highest-rated features in Figure~\ref{fig:prestudy} reflect these needs.
We grouped these features into four design goals (DG):
\begin{enumerate}
\item[\textbf{DG1:}] \textbf{Describe the surroundings, not only the route.}
Beyond turn-by-turn instructions, the system should tell the user what objects and obstacles are ahead, where they are, and how far away they are (F1, F2, F3).
\item[\textbf{DG2:}] \textbf{Localize the user reliably.}
The system should know the user's position and facing direction precisely enough so that its instructions remain valid at decision points, even where GPS and compass are weak (F4, F5).
\item[\textbf{DG3:}] \textbf{Keep the user's hands and ears free.}
Information should arrive automatically and hands-free, without occupying the cane hand or masking environmental sounds (F10).
The system should not require the user to carry a dedicated assistive device (F12).
\item[\textbf{DG4:}] \textbf{Speak in clock directions.}
The system should give both route guidance and object locations in clock directions relative to the user's facing direction (F11).
This format is more precise than left and right instructions, especially at complex intersections.
\end{enumerate}

\section{InSight}\label{sec:insight}

\begin{figure}[t!]
\centering
\includegraphics[width=0.9\textwidth]{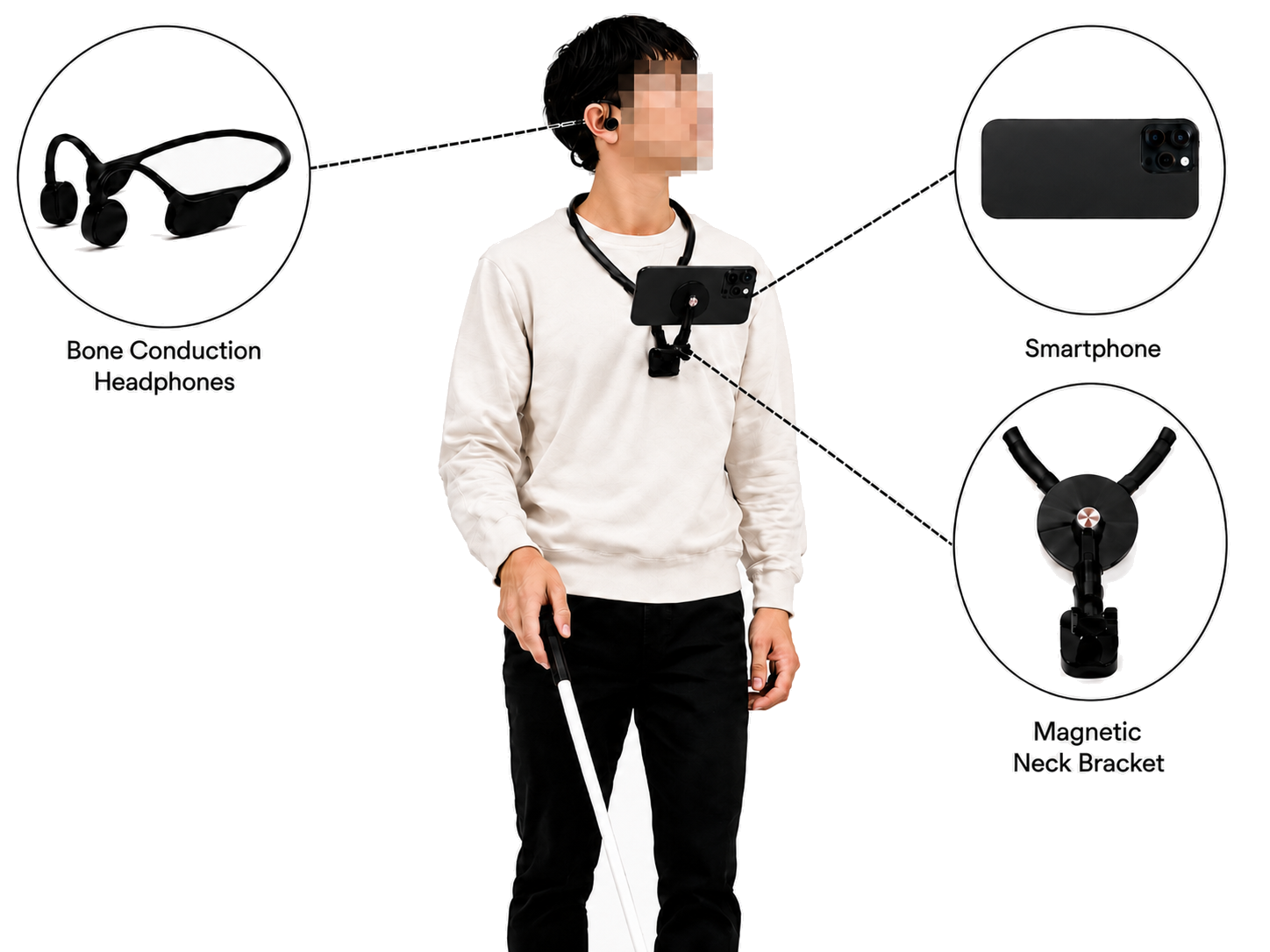}
\caption{Hardware setup of \textit{InSight}. The smartphone is mounted on the chest by a magnetic neck bracket, and speech is delivered through bone conduction headphones that leave both ears open. Both hands remain free, and the white cane stays in use.}\label{fig:concept}
\end{figure}

\textit{InSight} is designed as a mobile assistant.
A PVI pedestrian enters a destination, mounts the phone on the chest, and walks.
During the walk, the system reports two kinds of information through bone conduction headphones, namely route guidance toward the destination and descriptions of the objects ahead.
The design follows the simple principle that the user should not have to hold, aim, or query the phone while walking.
Figure~\ref{fig:concept} shows how \textit{InSight} is worn and used.
We structure the rest of this section as follows.
Section~\ref{sec:overview} provides an overview of the system (\textbf{what}).
Section~\ref{sec:mechanism} explains the mechanism of its two main functions (\textbf{how}).
Section~\ref{sec:rationale} presents the rationale behind the design decisions (\textbf{why}).
Section~\ref{sec:interface} describes the interface and interaction (\textbf{experience}).

\begin{figure}[t!]
\centering
\includegraphics[width=0.95\textwidth]{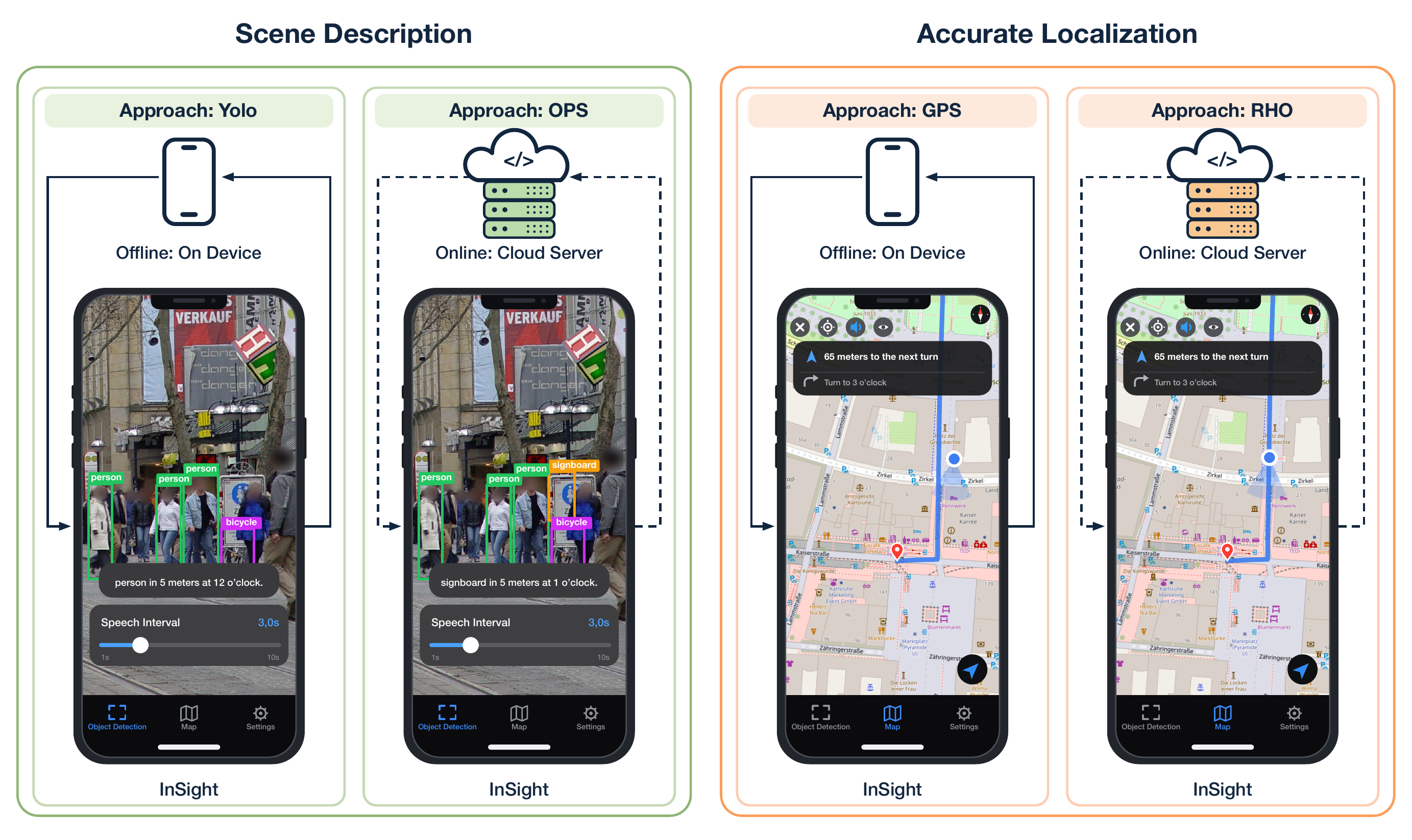}
\caption{The two main functions of \textit{InSight}: scene description (left) and accurate localization (right).
YOLO detects the people (green) and the parked bicycle (purple), whereas OPS also detects the signboard (orange), a category outside the fixed vocabulary.
GPS drifts off the planned route, whereas RHO stays on it.}\label{fig:architecture}
\end{figure}

\subsection{System Overview}\label{sec:overview}

\textit{InSight} implements the design goals as an iOS application with two operating modes.
Both modes share the same interaction design, the same announcement format, and the same settings, so switching between them changes what \textit{InSight} can recognize, not how it speaks.
They differ only in where object detection and camera localization are computed, and that choice is a trade-off.
Computing on the phone is fast and works without internet connection, but limits the system to a fixed category set and to GPS accuracy.
Computing on a server lifts both limits, but needs internet connection.
In the \textbf{offline mode} (\textit{InSight} Base), everything runs on the phone and no internet connection is required.
The on-device object detector YOLO11n~\cite{jocher2024yolo11} recognizes the 80 categories of the COCO dataset~\cite{lin2014microsoft} at 33 frames per second on the study device, an iPhone 12 Pro Max.
Localization relies on GPS and the compass.
In the \textbf{online mode} (\textit{InSight} Enhanced), the phone streams camera frames over a WebSocket connection to a self-hosted server.
The server runs two advanced models simultaneously.
OPS~\cite{zheng2024ops} is an open-vocabulary model and recognizes user-definable categories instead of a fixed list.
RHO~\cite{zheng2026rho} corrects the GPS position and compass heading against OSM data.
On its benchmark, RHO is able to place the camera within 3 meters of the true position in 74\% of the frames and within 5 degrees of the true heading in 90\% of the frames~\cite{zheng2026rho}.
In a city unseen during training, these rates still reach 67\% and 80\%.
Although OPS and RHO were originally developed for panoramic imagery, both models are compatible with the pinhole images captured by the phone camera.
The server returns about one frame every 3 seconds.
The online mode frees object detection from a fixed vocabulary and camera localization from GPS and compass errors.
The offline mode is the default, since it runs in real time and needs no internet connection.
The online mode gives richer descriptions and more accurate guidance, but the user must switch it on explicitly.
When the internet connection drops, \textit{InSight} falls back to the offline mode automatically so that guidance never stops.
Figure~\ref{fig:architecture} shows both functions and their data flows in the two modes.

\subsection{Main Functions}\label{sec:mechanism}

\subsubsection{Spoken Scene Descriptions}\label{sec:mech_scene}

\begin{figure}[t!]
\centering
\includegraphics[width=0.95\textwidth]{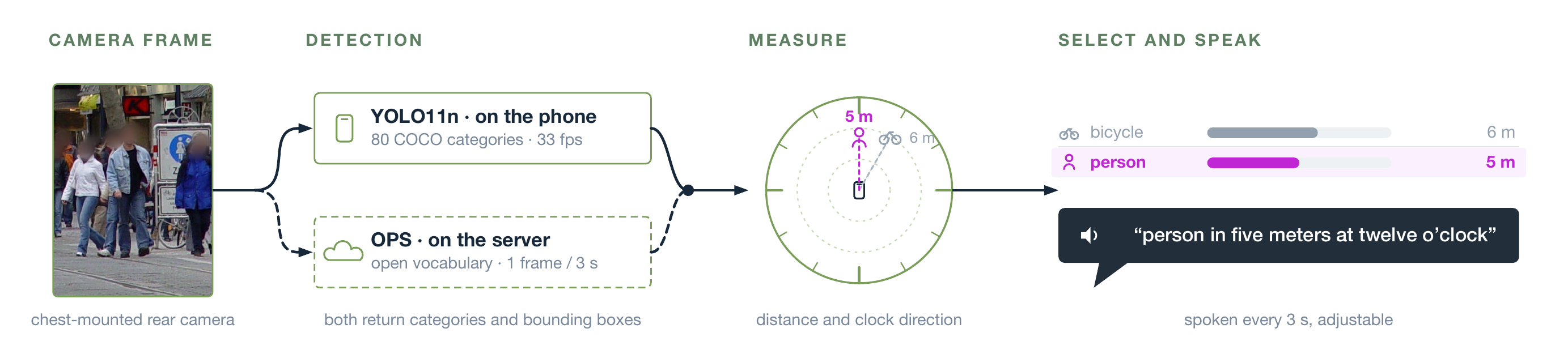}
\caption{The scene description pipeline.
A camera frame is processed by YOLO11n on the phone in the offline mode, or by OPS on the server in the online mode.
Both paths return categories and bounding boxes.
The phone then reads the distance from the depth map and converts the box position into a clock direction.
It filters the objects by the priority list and speaks the nearest eligible one.
The two modes differ only in the detection stage.}\label{fig:pipeline}
\end{figure}

The perception pipeline turns camera frames into short spoken announcements of the form ``\textit{[object] in [X] meters at [Y] o'clock}'' (DG1), for example, ``bicycle in two meters at eleven o'clock.''
Figure~\ref{fig:pipeline} shows the complete pipeline.
In the offline mode, frames from the rear camera are resized to $640\times640$ pixels and passed to YOLO11n~\cite{jocher2024yolo11}, a compact object detector deployed through Core ML.
Inference runs on the study device at 33 frames per second.
The distance to each detected object is read from ARKit's \texttt{sceneDepth} map, a dense depth image fused from the phone's LiDAR sensor and camera, sampled at the center of the detection box.
The horizontal position of the box is converted into a clock direction relative to the walking direction.
In the online mode, OPS outputs a semantic mask instead of object bounding boxes.
Connected-component labeling turns the mask into bounding boxes.
Each connected region of a category counts as one object.
The server sends the category and the bounding box of each object back to the phone.
The depth map stays on the phone, which samples \texttt{sceneDepth} at the returned bounding boxes exactly as in the offline mode, so both modes share the same distance and clock-direction calculation.

Announcing every detected object would overwhelm the user and distract from walking (F8, F9).
\textit{InSight} therefore announces selectively.
Users maintain an \emph{announcement priority list} that acts as an entry filter.
Only object categories on the list are eligible for announcement.
Among the eligible objects, the nearest one is announced.
By default the list contains seven categories.
Six are moving road users (person, bicycle, car, motorcycle, bus, truck) and the seventh is traffic light.
For traffic light the system announces only the location, not the signal state.
The crossing decision therefore stays with the user.
Announcements are spoken by the system's text-to-speech voice every 3 seconds by default, and the interval can be adjusted in the settings.
In the offline mode, users can extend the priority list with any of the 80 COCO categories~\cite{lin2014microsoft}.
In the online mode, since OPS is open-vocabulary, users can add arbitrary categories beyond COCO, such as signboards, bollards, or construction barriers, which no configuration of the offline detector could recognize.
On the server, an NVIDIA RTX PRO 6000 GPU processes about one frame every 3 seconds under good internet connection.

\subsubsection{Positioning and Route Guidance}\label{sec:mech_positioning}

\begin{figure}[t!]
\centering
\includegraphics[width=0.95\textwidth]{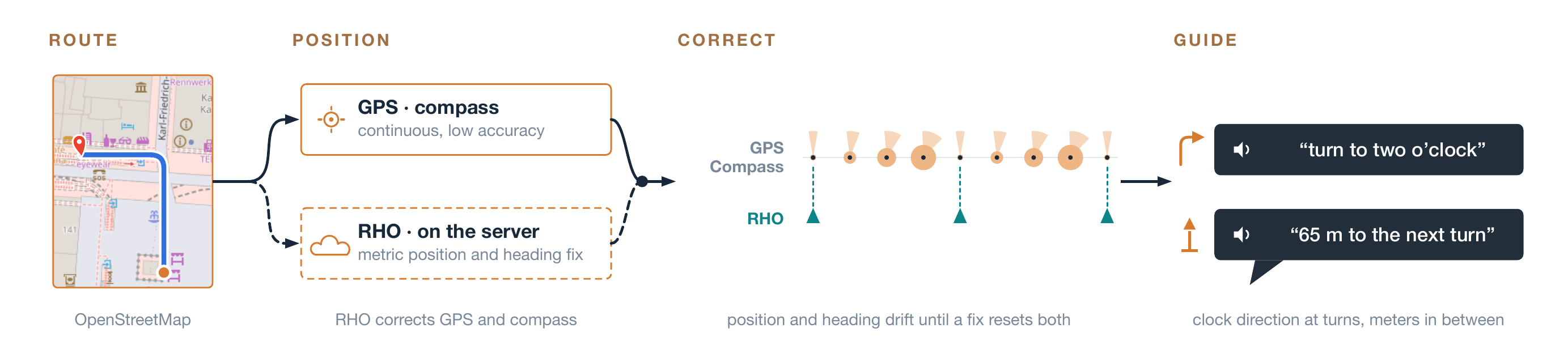}
\caption{Positioning and route guidance.
Routes are planned on OpenStreetMap data.
The offline mode locates the user from GPS and the compass.
In the online mode, RHO adds metric position and heading fixes that reset this error.
Guidance is spoken as a clock direction at turns and as a remaining distance on straight segments.}\label{fig:positioning}
\end{figure}

Route planning and guidance are based on OpenStreetMap in both modes.
Map data around the current GPS position is cached automatically, so \textit{InSight} works in any city without site-specific preparation.
Figure~\ref{fig:positioning} shows the positioning and guidance pipeline.
Guidance instructions are spoken continuously.
At turns, the system announces the new walking direction as a clock direction (``turn to two o'clock''), and on straight segments it announces the remaining distance in meters (``65 meters to the next turn'').

Positioning differs between the modes.
The offline mode relies on GPS and the compass, whose errors directly limit how well instructions align with the street.
The online mode adds RHO on top of this continuous but imprecise stream.
The server matches each frame against OSM data and returns a metric position and heading estimate.
This estimate then corrects the current GPS and compass readings (DG2).
The heading correction is equally important because a clock-direction instruction is defined relative to the user's heading, and any error in the compass reading shifts the announced direction by the same amount.

\subsection{Design Rationale}\label{sec:rationale}

The design of \textit{InSight} is the result of five deliberate decisions.
Each decision is justified below and traces back to the design goals and the findings of the pre-study.

\subsubsection{Clock Directions as the Single Direction Language}\label{sec:rat_clock}
All directional information, in scene descriptions and in route guidance, is expressed in clock directions (DG4).
The choice serves two purposes.
First, clock directions are familiar to PVI from orientation and mobility training~\cite{rocha2020om} and are finer-grained than ``left/right'', which lets the system express intermediate headings such as ``two o'clock'' at complex intersections.
Prior work on blind navigation adopted clock positions for the same reason~\cite{masaki2023pathfinder}.
Second, using one direction language for both information types gives the user a single mental model for the whole journey, instead of switching between ``slightly left'' for obstacles and street names for routes.
The pre-study supports this choice, as most respondents preferred clock directions over left and right instructions.

\subsubsection{One Object per Announcement}\label{sec:rat_oneobject}
Each announcement names exactly one object, the nearest eligible one.
This decision serves two purposes.
First, PVI pedestrians rely on ambient sound for orientation and hazard detection~\cite{kolarik2014echolocation}, so speech output must stay short to protect the auditory channel (DG3).
Second, the nearest object is the one that requires action first, so it is the appropriate choice when only one object can be announced.
The priority list keeps the announcement stream personally relevant, since which categories matter differs between users~\cite{bhanuka2023want,islam2024crucial}.

\subsubsection{Automatic Announcements with a User-Controlled Interval}\label{sec:rat_automatic}
\textit{InSight} delivers announcements actively rather than in response to a user query.
This decision serves two purposes.
First, query-based tools require the user to stop and operate the phone~\cite{BeMyAI2023,seeingAI}, which conflicts with hands-free walking (DG3).
Second, in an unfamiliar environment the user does not know what to ask about, so automatic announcements deliver exactly the information a query would have missed.
The default 3-second interval is adjustable, so the information rate stays under user control.
The pre-study supports this choice, as respondents wanted important objects announced automatically.

\subsubsection{Offline by Default, Online by Explicit Choice}\label{sec:rat_offline}
The dual-mode architecture serves two purposes.
First, it protects privacy.
In the default mode, camera images never leave the phone, and streaming them to the server requires an explicit choice that the user can revoke at any time.
Second, it keeps the system reliable.
A navigation aid must not fail with the network, so the offline mode is fully self-sufficient and the online mode falls back to it automatically when the connection drops.
This design reflects the pre-study, where attitudes toward processing camera images on a server were mixed.

\subsubsection{Chest Mount and Bone Conduction Headset}\label{sec:rat_mount}
A magnetic neck bracket holds the phone on the chest shown in Figure~\ref{fig:concept} and gives a stable, forward-facing camera view.
This mount serves two purposes.
First, the camera keeps a consistent viewpoint for detection and depth sampling.
Second, both hands stay free, so the white cane remains fully usable (DG3).
Speech is delivered through bluetooth bone conduction headphones that leave both ears open, preserving the environmental sounds that PVI depend on~\cite{he2023bone,kolarik2014echolocation}.
The pre-study supports this choice, as needing no dedicated assistive device was among the highest-rated needs (F12).

\subsection{Interface and Interaction}\label{sec:interface}

\begin{figure}[t!]
\centering
\includegraphics[width=\textwidth]{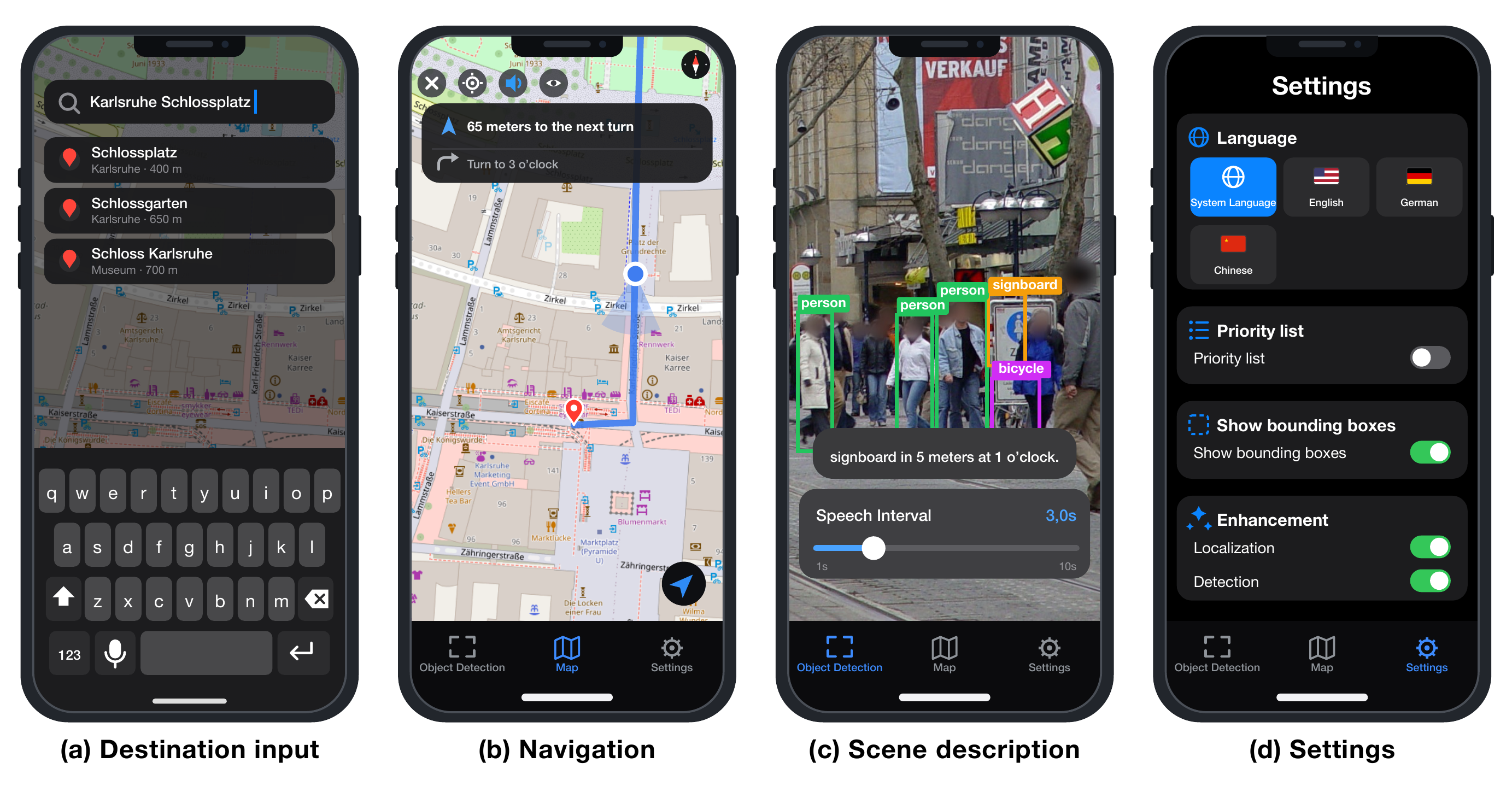}
\caption{The interface of \textit{InSight}.
(a) The user enters a destination by typing or by voice.
(b) During navigation, the map screen shows the planned route and the current guidance instruction.
(c) The scene description screen shows the camera view with the detected objects, the spoken announcement, and the slider that sets the announcement interval.
(d) The settings screen of \textit{InSight}.}\label{fig:interface}
\end{figure}

The user starts navigation by entering a destination, either by typing or by voice.
While walking, the user receives scene descriptions and route guidance as described in Section~\ref{sec:mechanism}, without holding or operating the phone.
The announcement interval is set on the scene description screen itself, so the user can change the speech rate without leaving the camera view.
The settings screen holds the speech language (English, German, and Chinese), the priority list, the display of bounding boxes, and the two switches that offload localization and detection to the server.
The entire interface is compatible with VoiceOver, the iOS screen reader that most blind iPhone users rely on~\cite{leporini2012voiceover}.
Figure~\ref{fig:interface} shows the four screens.

\section{User Study}\label{sec:study}

To answer RQ2 and RQ3, we conducted a within-subject field study with 21 PVI participants, comparing three conditions in unfamiliar outdoor areas: the commercial baseline Apple Maps, \textit{InSight} Base (offline mode), and \textit{InSight} Enhanced (online mode).
This study was approved by the Ethics Committee of Karlsruhe Institute of Technology (KIT), Germany.
The committee issued a formal ethics approval without assigning an approval number.

\subsection{Participants}\label{sec:participants}

We recruited 21 participants for the user study.
They were 13 men and 8 women ranging from 18 to 49 years old (\textit{M} = 29.1, \textit{SD} = 7.8).
Table~\ref{tab:participants} lists their profiles.
Vision level is defined based on the visual acuity of the better eye.
Under this definition, 10 participants had low vision, 8 had severe visual impairment, and 3 were blind.
For 9 participants the impairment had been present since birth, and the other 12 acquired it later in life.
All participants used a white cane as their primary mobility aid.
Participation was voluntary, and all participants provided informed consent for the study and for video and audio recording.

\begin{table}[t!]
\tbl{Profiles of the 21 participants of the user study.
Blind: a visual acuity of 2\% or less.
Severe impairment: above 2\% up to 5\%.
Low vision: above 5\% up to 30\%.}
{\setlength{\tabcolsep}{4pt}
\begin{tabular*}{\textwidth}{@{\extracolsep{\fill}}lllllll@{}} \\[-5.5pt]\noalign{\hrule height .8pt}\\[-5pt]
ID & Gender & Age group & Vision level & Onset & Mobility aids & Assistive apps \\ \midrule
P1 & Male & 20--29 & Severe impairment & Later in life & White cane & Apple Maps, Be My AI \\
P2 & Female & 20--29 & Low vision & Later in life & White cane & Apple Maps \\
P3 & Female & 30--39 & Severe impairment & Later in life & White cane & Apple Maps, Be My AI \\
P4 & Female & 30--39 & Low vision & Later in life & White cane & Be My AI \\
P5 & Female & 30--39 & Severe impairment & Since birth & White cane & Apple Maps, Be My AI \\
P6 & Male & 30--39 & Blind & Later in life & White cane & BlindSquare \\
P7 & Female & 20--29 & Low vision & Later in life & White cane & Apple Maps \\
P8 & Female & 20--29 & Blind & Since birth & White cane & Apple Maps, Be My AI \\
P9 & Male & 20--29 & Low vision & Later in life & White cane & Apple Maps, BlindSquare \\
P10 & Male & 20--29 & Low vision & Since birth & White cane & Be My AI \\
P11 & Male & 30--39 & Severe impairment & Later in life & White cane & Apple Maps, BlindSquare \\
P12 & Male & 18--19 & Low vision & Later in life & White cane & Apple Maps, Be My AI \\
P13 & Male & 20--29 & Severe impairment & Since birth & White cane & BlindSquare, Be My AI \\
P14 & Male & 18--19 & Severe impairment & Since birth & White cane & Apple Maps \\
P15 & Female & 20--29 & Low vision & Since birth & White cane & BlindSquare, Be My AI \\
P16 & Male & 20--29 & Severe impairment & Later in life & White cane & BlindSquare, Be My AI \\
P17 & Male & 30--39 & Blind & Since birth & White cane & Apple Maps, BlindSquare \\
P18 & Male & 40--49 & Low vision & Since birth & White cane & BlindSquare, Be My AI \\
P19 & Male & 30--39 & Low vision & Later in life & White cane & Apple Maps, BlindSquare \\
P20 & Female & 30--39 & Severe impairment & Since birth & White cane & Apple Maps, Be My AI \\
P21 & Male & 20--29 & Low vision & Later in life & White cane & BlindSquare \\ \\[-7.5pt]\noalign{\hrule height .8pt}
\end{tabular*}}
\label{tab:participants}
\end{table}

\subsection{Procedure}\label{sec:procedure}

\begin{figure}[t!]
\centering
\includegraphics[width=\textwidth]{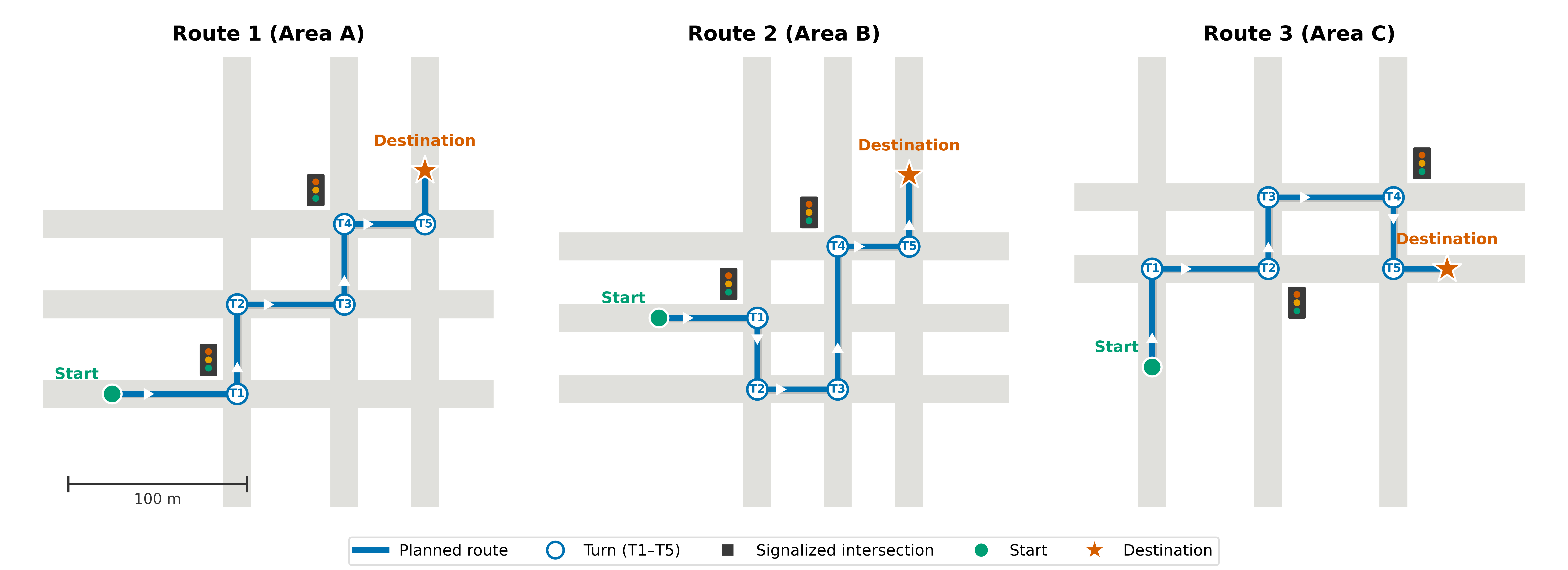}
\caption{Schematic layout of the three navigation routes used in the main study. Each route was about 300 meters long containing five turns (T1--T5) and two signalized intersections, so that the three routes were matched in length and complexity. Each route lay in a separate area that no participant had visited before.}\label{fig:routes}
\end{figure}

Each session lasted about two hours.
After the welcome and consent procedure, participants completed a background questionnaire and then an approximately 10-minute familiarization session on a separate training route in a distinct area, where they tried the systems and the chest mount.
The main part consisted of three navigation tasks, one per condition.
Each of the three routes was approximately 300 meters long with five turns and two signalized intersections, located in representative everyday areas (office buildings, street intersections, pedestrian pathways).
The three routes were matched in length and complexity, and none of the participants had ever visited any of the three areas.
Figure~\ref{fig:routes} shows the three routes, each in a separate area (Areas A to C).
Each condition was bound to one area so that no route knowledge could transfer between conditions.
In the baseline condition, participants navigated with Apple Maps' standard spoken pedestrian guidance.
In the two \textit{InSight} conditions, participants wore the chest-mounted phone and used the same interface and announcement format, so that these two conditions differed only in the operating mode described in Section~\ref{sec:overview}.
\textit{InSight} Base ran entirely on the phone, whereas \textit{InSight} Enhanced streamed camera frames to the server for open-vocabulary recognition and for visual correction of position and heading.
The order of the three conditions was counterbalanced across participants with a $3 \times 3$ Latin square repeated seven times, so that each of the three orders was assigned to seven participants.
The white cane was available in all conditions during navigation.

Two experimenters accompanied every trial: one walked close to the participant exclusively as a safety guard and intervened only when the participant's physical safety was at risk; the other recorded video and experimental data.
Safety interventions did not count as help requests.
Participants were encouraged, but not required, to think aloud.
After each task, participants filled in two questionnaires, namely the NASA Task Load Index (NASA-TLX) for perceived workload and the System Usability Scale (SUS) for perceived usability.
After the last task, a semi-structured interview concluded the session.
All sessions were audio- and video-recorded and anonymized.

\subsection{Measures}\label{sec:measures}

For RQ2, we measured six task performance indicators per trial:
\begin{itemize}
\item \emph{Success} was recorded when the participant finished within 5 m of the destination.
\item \emph{Final distance to goal} is the distance between the stopping point and the destination.
\item \emph{Time to reach} is the duration from start to stop.
\item \emph{Collisions} counts unintended contacts of the participant's body with obstacles or persons. Cane contacts were not counted, as probing is the cane's intended function.
\item \emph{Help requests} counts the times the participant actively asked the experimenters for help.
\item \emph{Wrong turns} counts the decision points (turns or intersections) at which the participant set off in a direction other than the planned one.
\end{itemize}

For RQ3, participants completed two standard questionnaires after each condition.
The NASA Task Load Index~\cite{hart2006nasa} measured perceived workload on six dimensions (mental demand, physical demand, temporal demand, performance, effort, frustration), each rated on the standard 21-point graphical scale and converted to a score from 0 to 100 in increments of 5.
We did not collect the pairwise dimension weights.
We therefore report the raw TLX score (RTLX), the unweighted mean of the six dimension ratings.
RTLX ranges from 0 to 100, and higher values indicate higher perceived workload.
The System Usability Scale~\cite{brooke1996sus} measured perceived usability with ten items, scored 0 to 100 by the standard formula.
Qualitative data on the reasons behind the ratings were collected in the closing semi-structured interviews.

\section{Conclusion}\label{sec:conclusion}

In this article, we presented \textit{InSight}, an iOS navigation assistant for pedestrians with visual impairments that integrates spoken scene descriptions and precise positioning, offline on the phone by default and enhanced by open-vocabulary segmentation and OpenStreetMap-based localization when the user opts in.
Grounded in a formative pre-study and evaluated in a within-subject field study with 21 PVI participants in unfamiliar outdoor areas, \textit{InSight} outperformed a commercial baseline on every task performance, workload, and usability measure, with the enhanced mode reaching an 86\% success rate and a SUS score of 75.
The findings show that a phone PVI already carry can give back much of what unfamiliar environments take away: awareness of the surroundings and certainty about position and heading.
This article aims to inspire the accessibility community to treat scene understanding and positioning not as separate features but as one integrated, user-controlled information stream for independent travel.

\bibliographystyle{apacite}
\bibliography{references}

\end{document}